\documentclass[aps,preprint,groupedaddress]{revtex4-1}
\usepackage{amssymb}
\usepackage{mathrsfs}
\usepackage{slashed}
\usepackage[breaklinks]{hyperref}
\usepackage{graphicx}
\usepackage{subfigure}

\def\figurewidth{0.4\textwidth}
\def\minipagewidth{0.4\textwidth}
\def\subfigurewidth{1\textwidth}

\begin{document}

\title{QCD factorization for $\Lambda_b^0\to\Lambda_c^+\pi^-$}

\author{Zhen-Hua Zhang}
 \email[Email: ]{zhangzh@mail.bnu.edu.cn}
\author{Xin-Heng Guo}%
 \email[\textit{Corresponding author}, Email: ]{xhguo@bnu.edu.cn}
\affiliation{%
 College of Nuclear Science and Technology, Beijing Normal University, Beijing 100875, China
}%
\author{Gang L\"{u}}
 \email[Email: ]{ganglv@haut.edu.cn}
\affiliation{
 College of Science, Henan University of Technology, Zhengzhou 450001, China.}

\date{\today}

\begin{abstract}
We prove that in the limit $m_b, m_c\to\infty$ with $m_c/m_b$ fixed, factorization holds at order $\alpha_s$ for the decay $\Lambda_b^0\to\Lambda_c^+\pi^-$.
This proof is done in the infinite-momentum frame in which the momenta of  $\pi$, $\Lambda_c$, and $\Lambda_b$ go to infinity.
Our result is renormalization-scale- and scheme-independent at $\mathcal{O}(\alpha_s)$.
This is the same as the QCD factorization for $B \to D \pi$.
\end{abstract}


\maketitle

%\section*{\label{introduction}Introduction.}

Nonleptonic weak decays of heavy hadrons are of great interest, since they provide a good area to understand the standard model and to find new physics beyond the standard model.
For these decays, the evaluation of the hadronic matrix elements is the most difficult problem due to the involvement of nonperturbative QCD effects that prevent us from the complete treatment of these matrix elements within the perturbative approach.
QCD factorization was proven to be applicable to b-meson decays by Beneke, Buchalla, Neubert, and Sachrajda \cite{Beneke:1999br,Beneke:2000ry,Beneke:2001ev}.
They showed that in the case of heavy-light final states only factorizable diagrams contribute in the heavy-quark limit, leading to easier treatment of the hadronic matrix elements.

There have been far fewer theoretical studies for heavy baryons than for heavy mesons.
In recent years, more and more experimental results about heavy baryons have been reported by various experimental collaborations \cite{VanKooten:2010zz,Acosta:2005mq}.
In the near future, the copious production of heavy baryons is expected at the Large Hadron Collider at CERN.
Therefore, it is urgent to study heavy baryons in much more detail theoretically.
It is the aim of this paper to study heavy-baryon nonleptonic decays in the framework of QCD factorization, focusing on the decay $\Lambda_b^0\to\Lambda_c^+\pi^-$.
This situation is much more complex than the B meson decays because the additional light quark in $\Lambda_b$ (and $\Lambda_c$) generates many more Feynman diagrams.
However, we will prove that most of these diagrams are power-suppressed in the heavy-quark limit, leaving only the factorizable vertex corrections at $\mathcal{O}(\alpha_s)$.
This means that factorization holds at $\mathcal{O}(\alpha_s)$ for the decay $\Lambda_b^0\to\Lambda_c^+\pi^-$.

The effective Hamiltonian for the weak decay $\Lambda_b^0\to\Lambda_c^+\pi^-$ is \cite{Buchalla:1995vs}
\begin{equation}
\mathscr{H}_{\mathrm{eff}}=\frac{G_F}{\sqrt{2}}V_{ud}^*V_{cb}\left[c_1(\mu)Q_1+c_2(\mu)Q_2\right]+H.c.,\label{hamilton}
\end{equation}
where $G_F$ is the Fermi constant, $c_1$ and $c_2$ are Wilson coefficients at the scale $\mu$ [which is $\mathcal{O}(m_b)$], $V_{ud}$ and $V_{cb}$ are Cabibbo-Kobayashi-Maskawa matrix elements, and $Q_1$ and $Q_2$ are four-quark operators that have the following form:
\begin{eqnarray}
&&Q_1=\bar{d}\gamma_{\mu}(1-\gamma_5)u\bar{c}\gamma^{\mu}(1-\gamma_5)b, \nonumber\\
&&Q_2=\bar{d}^{i}\gamma_{\mu}(1-\gamma_5)u^{j}\bar{c}^{j}\gamma^{\mu}(1-\gamma_5)b^{i},
\end{eqnarray}
where $u$, $d$, $c$, and $b$ represent quark-field operators, and the superscripts $i$ and $j$ are color indices.

The typical radii of baryons are of the order $1/\Lambda_{QCD}$ because of QCD confinement.
This allows us to make the following power-counting rule for the wave functions of the valence Fock state $\Psi_{X}$ ($X$ stands for $\pi$, $\Lambda_c$, $\Lambda_b$):
it is of the order $f_X\Phi_X/\Lambda_{QCD}^{\delta_X}$ ($\delta_{X}=2$ for $X=\pi$, and $\delta_{X}=4$ for $X=\Lambda_b,\Lambda_c$) when the transverse momenta of the valence quarks $\sim \Lambda_{QCD}$ and 0 when the transverse momentum of at least of one of the valence quarks $\gg \Lambda_{QCD}$, where $f_X$ is the decay constant of $X$, and $\Phi_X$ is the light-cone distribution amplitude of $X$.
Following Ref. \cite{Beneke:2000ry}, the light-cone distribution amplitude $\Phi_{\pi}$ has the following power-counting rule:
$\Phi_{\pi}$ is $\mathcal{O}(1)$ when the longitudinal-momentum fractions of the valence quarks are $\mathcal{O}(1)$, while $\Phi_{\pi}$ is $\mathcal{O}(\Lambda_{QCD}/m_b)$ when either of the longitudinal-momentum fractions of the valence quarks is $\mathcal{O}(\Lambda_{QCD}/m_b)$.
Similar to B mesons, we find that for $\Lambda_{Q}$ ($Q$ stands for $b$ or $c$), $\Phi_{\Lambda_Q}$ is of $\mathcal{O}[(m_{\Lambda_Q}/\Lambda_{QCD})^2]$ when both the longitudinal momentum fractions of the light quarks are  of $\mathcal{O}(\Lambda_{QCD}/m_{\Lambda_Q})$ and is 0 elsewhere.
The power-counting rules for the decay constants are
\begin{equation}
f_{\pi}\sim\Lambda_{QCD}, ~~~
f_{\Lambda_Q}\sim \frac{\Lambda_{QCD}^3}{m_{\Lambda_Q}},
\end{equation}
which follow directly from the power-counting rules of the wave functions and the light-cone distribution amplitudes and their normalization conditions.

Our work is done in the infinite-momentum frame of $\Lambda_b$.
We start with the rest frame of $\Lambda_b$, in which $\pi$ moves along the z axis; then, we go to the infinite-momentum frame by making a Lorentz boost along the z axis.
In this frame, the energy of $\Lambda_b$ (denoted by $E$), the mass of $\Lambda_b$, and the QCD scale $\Lambda_{QCD}$ are hierarchically ordered as  $E \gg m_{\Lambda_b} \gg \Lambda_{QCD}$.
The relations of the momenta of the three particles are $p_{\Lambda_b}:p_{\Lambda_c}:p_{\pi}=1:z^2:(1-z^2)$, where $z=m_c/m_b$ (in the heavy-quark limit $m_{\Lambda_c}/m_{\Lambda_b}=m_c/m_b$).

\begin{figure}
\includegraphics[width=\figurewidth]{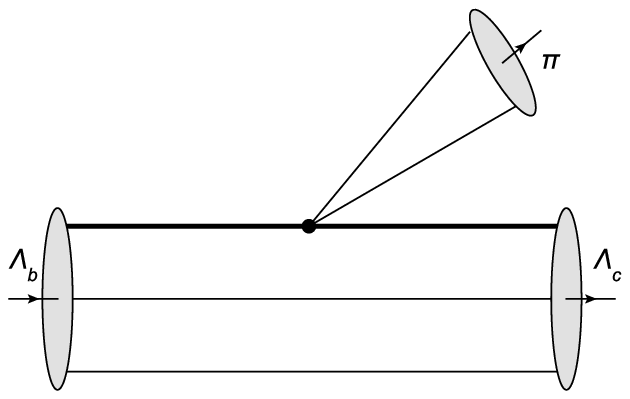}\\
\caption{\label{tree1}The factorizable diagram at the tree level, denoted by $T_A$. Heavy lines represent heavy quarks (same in the following figures).}
\end{figure}

\begin{figure}
\includegraphics[width=\figurewidth]{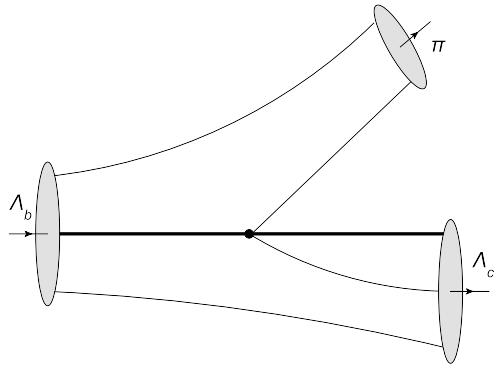}\\
\caption{\label{tree2} The nonfactorizable diagram at the tree level, denoted by $T_B$.}
\end{figure}

The diagrams in Figs. \ref{tree1} and \ref{tree2} are tree diagrams (denoted by $T_A$ and $T_B$, respectively).
Figure \ref{tree1} is factorizable.
The probability of finding a hadron in its valence Fock state is $\mathcal{O}(1)$.
We use the valence Fock states of $\Lambda_b$, $\Lambda_c$, and $\pi$ to make the power estimation (we will show later that higher-Fock-state contributions are either power-suppressed or can be absorbed into the form factors of factorizable diagrams).
First of all, $T_A$ has the order of the following:
\begin{equation}
T_A \sim \langle\pi^{-}|\bar{d}\gamma_{\mu}(1-\gamma_5)u|0\rangle \cdot\langle\Lambda_c^+|\bar{c}\gamma^{\mu}(1-\gamma_5)b|\Lambda_b^0\rangle.
\end{equation}
Note that the above expression holds for both $Q_1$ and $Q_2$ in Eq. (\ref{hamilton}). The difference between their contributions is just a color factor which does not change the order.
The two matrix elements in the above expression are related to the decay constant of $\pi^-$ and the form factors of the transition $\Lambda_b^0 \to \Lambda_c^+$, respectively.
In the heavy-quark limit, one has \cite{Mannel:1990vg,Falk:1992ws}
\begin{equation}
\langle\Lambda_c|\bar{c}\gamma^{\mu}(1-\gamma_5)b|\Lambda_b\rangle=\zeta(\omega)\bar{u}(p_{\Lambda_c})\gamma^{\mu}(1-\gamma_5)u(p_{\Lambda_b}),
\end{equation}
where $\zeta(\omega)$ is the Isgur-Wise function ($\omega$ is the velocity transfer).
Then, it can be seen that $T_A$ is of the order $ f_{\pi}\bar{u}(p_{\Lambda_c})\slashed{p}_{\pi}(1-\gamma_{5})u(p_{\Lambda_b})$.
The diagram in Fig. \ref{tree2} would violate factorization, but, in the following, we will prove that this diagram is power-suppressed compared to that in Fig. \ref{tree1} in the heavy-quark limit.

From the structure of Fig. \ref{tree2}, one can see that $T_B$ can be expressed as an overlap of wave functions of $\Lambda_b^0$, $\Lambda_c^+$, and $\pi^-$ as the following:
\begin{eqnarray}
T_{B} &\sim & \int [d\xi dp_{\perp}] \Psi_{\pi}^{*}\Psi_{\Lambda_c}^{*}\Psi_{\Lambda_b} \bar{u}(p_{c_{\Lambda_c}})\gamma^{\mu}(1-\gamma_5)u(p_{b_{\Lambda_b}}) E_{d_{\pi}}\delta^{(3)}(\vec{p}_{d_{\pi}}-\vec{p}_{d_{\Lambda_b}}) E_{u_{\Lambda_b}}\delta^{(3)}(\vec{p}_{u_{\Lambda_c}}-\vec{p}_{u_{\Lambda_b}})\nonumber\\ && \times C_{r_1r_2}\bar{u}_d^{r_1}(p_{d_{\Lambda_c}})\gamma_{\mu}(1-\gamma_5)v_u^{r_2}(p_{u_{\pi}}),
\end{eqnarray}
where the subscripts $q_X$ ($q=u, d, c, b$, $X=\Lambda_b^0, \Lambda_c^+, \pi^-$) represent the quark $q$ in the hadron $X$;
$C_{r_1r_2}$ ($r_1$ and $r_2$ are helicity indices with values of 1 or 2) are the Clebsch-Gordan coefficients that equal $(-)1/\sqrt{2}$ for $r_1=r_2=1(2)$ and $0$ elsewhere; and
\begin{eqnarray}
[d\xi dp_{\perp}]&=&\prod_{X=\Lambda_b^0, \Lambda_c^+, \pi^-} \Bigg[\left(\prod_{q}\frac{d\xi_{q_{X}}d^2p_{q_{X}\perp}}{2(2\pi)^3\sqrt{\xi_{q_{X}}}}\right) 2(2\pi)^3\delta\left(1-\sum_q\xi_{q_{X}}\right) \delta^{(2)}\left(\sum_{q}\vec{q}_{q_{X}\perp}\right)\Bigg],
\end{eqnarray}
with $\xi_{q_{X}}$ and $p_{q_{X}\perp}$ representing the longitudinal-momentum fraction and the transverse momentum of the quark $q$ in the hadron $X$, respectively.
Since we are working in the infinite-momentum frame, we can replace the spinor $u(p_{q_{X}})$ by $\sqrt{\xi_{q_{X}}}u(p_{X})$ safely.
Taking into account the power-counting rules for the valence Fock-state wave functions, we find that $T_B$ is of the order
\begin{eqnarray}
&&\frac{1}{\Lambda_{QCD}^4}f_{\pi}f_{\Lambda_c}f_{\Lambda_b}\bar{u}(p_{\Lambda_c})\gamma^{\mu}(1-\gamma_5)u(p_{\Lambda_b})\cdot C_{r_1r_2} \bar{u}^{r_1}(p_{\Lambda_c})\gamma_{\mu}(1-\gamma_5)v^{r_2}(p_{\pi}).
\end{eqnarray}
We can express the spinor $u(p)$ of a baryon (with mass $m\neq 0$) by a Lorentz boost from its rest frame: $u(p)=\Lambda_{\frac{1}{2}}(p,m)u(m)$, where $\Lambda_{\frac{1}{2}}(p,m)$ is the spinor representation of the Lorentz transformation, which boosts a four-vector $(m,\vec{0})$ to $p$.
Then, after some straightforward derivation, one can obtain the order of $T_B$, which is
\begin{eqnarray}
&&\frac{1}{\Lambda_{QCD}^4}f_{\pi}f_{\Lambda_c}f_{\Lambda_b}\bar{u}(p_{\Lambda_c})\gamma^{\mu}(1-\gamma_5)u(p_{\Lambda_b}) \frac{1}{\sqrt{m_{\Lambda_c}m_{\pi}}} \nonumber\\ && \times\mathrm{tr}\big[\gamma_{\mu}(1-\gamma_5) (\slashed{p}_{\pi}-m_{\pi})\Lambda_{\frac{1}{2}}(p_{\pi},m_{\pi})\gamma_5\Lambda_{\frac{1}{2}}^{-1}(p_{\Lambda_{c}},m_{\Lambda_c}) (\slashed{p}_{\Lambda_{c}}+m_{\Lambda_c})\big]. \label{TB order}
\end{eqnarray}
We find that the factor $\Lambda_{\frac{1}{2}}(p_{\pi},m_{\pi})\gamma_5\Lambda^{-1}_{\frac{1}{2}}(p_{\Lambda_{c}},m_{\Lambda_c})$ has the form of
\begin{equation}
\frac{1}{m_{\Lambda_b}^2\sqrt{z(1-z^2)}}\sqrt{\frac{m_{\Lambda_b}}{m_{\pi}}}\gamma_5 \Big[\slashed{p}_{\pi}\slashed{p}_{\Lambda_c}+\frac{zm_{\pi}}{(1-z^2)m_{\Lambda_b}}\slashed{p}_{\Lambda_c}\slashed{p}_{\pi}\Big], \label{GgG}
\end{equation}
in the frame in which we are working.
The leading term $\slashed{p}_{\pi}\slashed{p}_{\Lambda_c}$ is cancelled when we substitute Eq. (\ref{GgG}) into Eq. (\ref{TB order}), so the order of $T_B$ is $T_B\sim T_A (\Lambda_{QCD}/m_b)^2$.
Therefore, the only diagram that contributes at the leading order of $\alpha_s$ is the one in Fig. \ref{tree1} in the heavy-quark limit, which is factorizable.
In this situation, the decay amplitude for $\Lambda_b^0\to\Lambda_c^{+}\pi^{-}$ can be written in the form of the so-called naive factorization (NF),
\begin{equation}
\mathcal{A}_{NF}
=ia_1(\mu)\frac{G_{F}}{\sqrt{2}}V_{ud}^*V_{cb}f_{\pi}\zeta(\omega)\bar{u}(p_{\Lambda_c})\slashed{p}_{\pi}(1-\gamma_5)u(p_{\Lambda_c}), \label{NF decay amplitude}
\end{equation}
where $a_1(\mu)=c_1(\mu)+c_2(\mu)/N_c$ ($N_c$ is the color number).
We can reexpress $a_{1}(\mu)$ as
\begin{eqnarray}
a_{1}(\mu)&=&\bar{c}_1(m_b)+\frac{\bar{c}_2(m_b)}{N_c}\bigg[1+\frac{\alpha_s(\mu)}{4\pi}C_F\Big(11+\kappa_{+}+6\ln\frac{\mu^2}{m_b^2}\Big)\bigg],
\end{eqnarray}
where $\bar{c}_1(m_b)$ and $\bar{c}_2(m_b)$ are scheme-independent Wilson coefficients, $C_F=(N_c^2-1)/(2N_c)$, and $\kappa_+$ is a scheme dependent parameter that equals $0$ ($\pm4$) in the Nave Dimensional Regularization (t' Hooft-Veltman) scheme.
We can see that the scheme and scale dependence of $a_{1}(\mu)$ is an $\mathcal{O}(\alpha_s)$ effect.

Next, we will prove that at order $\alpha_s$, factorization still holds in the heavy-quark limit.
At $\mathcal{O}(\alpha_s)$, the diagrams can be classified into three categories: corrections to $T_A$, corrections to $T_B$, and annihilation diagrams.
We will discuss each of them in the following.

\begin{figure}
\centering
\subfigure{
\begin{minipage}[b]{\minipagewidth}
\centering
\includegraphics[width=\subfigurewidth]{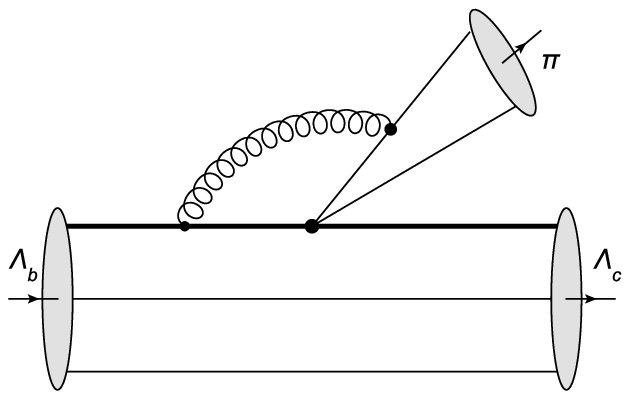}
\end{minipage}
}%
\hspace{0.04\linewidth}
\subfigure{
\begin{minipage}[b]{\minipagewidth}
\centering
\includegraphics[width=\subfigurewidth]{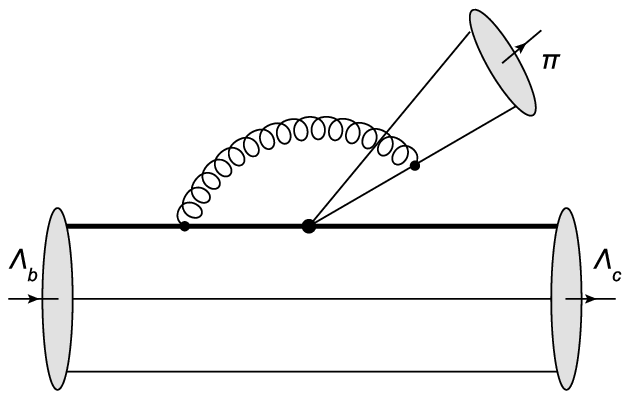}
\end{minipage}
}\\
\subfigure{
\begin{minipage}[b]{\minipagewidth}
\centering
\includegraphics[width=\subfigurewidth]{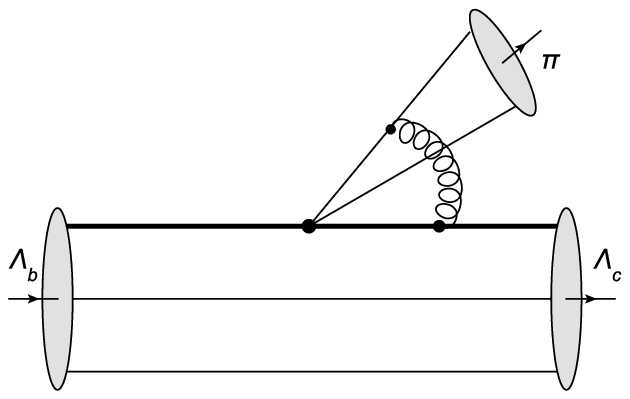}
\end{minipage}
}%
\hspace{0.04\linewidth}
\subfigure{
\begin{minipage}[b]{\minipagewidth}
\centering
\includegraphics[width=\subfigurewidth]{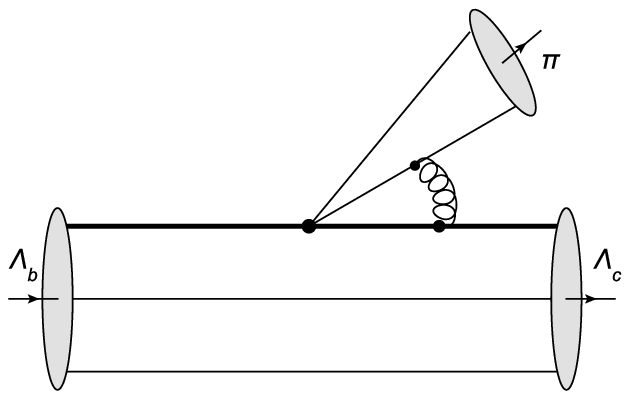}
\end{minipage}
}%
\caption{\label{vertexcorrection}Vertex corrections to $T_A$.}
\end{figure}

There are two kinds of diagram that are corrections to $T_A$: one kind is the vertex corrections shown in Fig. \ref{vertexcorrection}, which are factorizable, and the other kind is the so-called ``nonfactorizable'' spectator-scattering diagrams similar to those in Ref. \cite{Beneke:2000ry}.
Unlike two nonfactorizable spectator-scattering diagrams in the case of $B \to D\pi$, there are four such diagrams for $\Lambda_b^0 \to \Lambda_c^+ \pi^-$ due to an additional light quark in $\Lambda_b$ (and $\Lambda_c$).
One can find that each of them are of the order $\alpha_sT_A$, which is not suppressed.
However, the leading terms cancel when one sums up all of the four diagrams, leading to power suppression.
This is similar to the case of QCD factorization for $B \to D\pi$\cite{Beneke:2000ry}.

There are, in total, 11 diagrams for the $\mathcal{O}(\alpha_s)$ corrections to $T_B$, which can be classified into two categories.
The first category is the vertex corrections to $T_B$ that contain one loop, while the second category is tree diagrams.
We prove that all of them are power-suppressed.
The first kind of diagram contains infrared and ultraviolet divergences because of the loops, but since this kind of diagram is of the order $\alpha_s T_B$, these diagrams are still power-suppressed in the heavy-quark limit.

\begin{figure}
\centering
\subfigure{
\begin{minipage}[b]{\minipagewidth}
\centering
\includegraphics[width=\subfigurewidth]{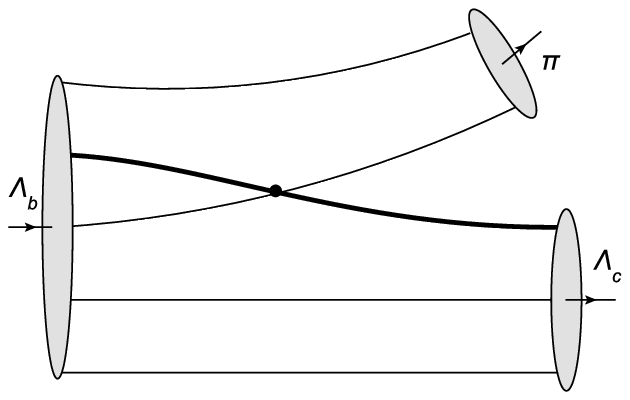}
\end{minipage}
}%
\hspace{0.04\linewidth}
\subfigure{
\begin{minipage}[b]{\minipagewidth}
\centering
\includegraphics[width=\subfigurewidth]{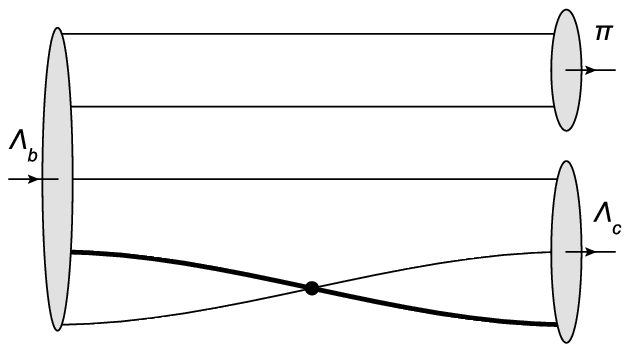}
\end{minipage}
}
\caption{\label{highfockstate}Two examples for $\Lambda_b \to \Lambda_c\pi$ when $\Lambda_b$ and/or $\Lambda_c$ are at their nonvalence Fock states.}
\end{figure}

Now, we discuss the contributions from nonvalence Fock states.
The proof for the suppression of higher-Fock-state contributions of the $\pi$ meson is very similar to the decay $B\to D\pi$ \cite{Beneke:2000ry}.
We still have to consider the situation when $\Lambda_b$  and/or $\Lambda_c$ are at their nonvalence Fock states while $\pi$ is at its valence Fock state.
Most of this kind of diagram are factorizable and can be absorbed into the form factor of the transition $\Lambda_b\to\Lambda_c$.
However, when (at least) one of the valence quarks of $\pi$ comes form the sea quarks of $\Lambda_b$, the diagram is nonfactorizable and cannot be absorbed into the form factor.
We have to consider this kind of diagram explicitly.
Figure \ref{highfockstate} shows two typical diagrams of this kind.
To prove the suppression of these diagrams, we have to give the power-counting rules for the nonvalence Fock states of $\Lambda_b$.
Notice that the probability that $\Lambda_b$ is in its valence Fock state is $\mathcal{O}(1)$; the power of the wave function of the nonvalence Fock state is, at most, the order obtained, if we assume that the probability that $\Lambda_b$ is in the nonvalence Fock state is a constant (that is not suppressed by $1/m_b$).
This means that we can adopt similar power-counting rules to that of the valence Fock state.
The wave functions of the $n$-parton nonvalence Fock state $\Psi^{(n)}_{\Lambda_b}$ are of the order $f^{(n)}_{\Lambda_b}\Phi_{\Lambda_b}^{(n)}/\Lambda_{QCD}^{2n-2}$ when the transverse momenta of all partons $\sim \Lambda_{QCD}$; and are 0 when the transverse momentum of at least one of the partons $\gg \Lambda_{QCD}$.
The light-cone distribution amplitude is $\Phi_{\Lambda_b}^{(n)}\sim(m_{\Lambda_b}/\Lambda_{QCD})^{n-1}$ when all the fractions of the longitudinal momenta of the light partons are of $\mathcal{O}(\Lambda_{QCD}/m_{\Lambda_Q})$ and is 0 elsewhere.
The power-counting rule for $f^{(n)}_{\Lambda_b}$ is $\Lambda_{QCD}^{\frac{3}{2}(n-1)}/m_{\Lambda_b}^{\frac{1}{2}(n-1)}$.
With the above power-counting rules, we find that the first diagram of Fig. \ref{highfockstate} is power-suppressed.
The second diagram is even more suppressed, because at least one of the partons that goes into $\pi$ is collinear, \textit{i.e.}, the fraction of its momentum is $\mathcal{O}(1)$; however, this cannot happen because of the power-counting rule of the light-cone distribution amplitude of $\Lambda_b$.
There may be other diagrams with even more partons in $\Lambda_b$ and $\Lambda_c$.
But with the restriction that $\pi$ should be at its valence Fock state, the extra partons should go directly from $\Lambda_b$ to $\Lambda_c$.
This situation is similar to either the first or the second diagram in Fig. \ref{highfockstate}.
Then, we complete the proof of suppression of nonvalence Fock-state diagrams.

\begin{figure}
\includegraphics[width=\figurewidth]{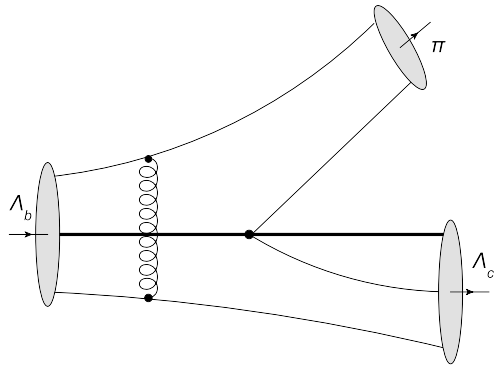}\\
\caption{\label{example1}An example diagram that should not be considered. This diagram in fact belongs to $T_B$.}
\end{figure}

\begin{figure}
\centering
\subfigure{
\begin{minipage}[b]{\minipagewidth}
\centering
\includegraphics[width=\subfigurewidth]{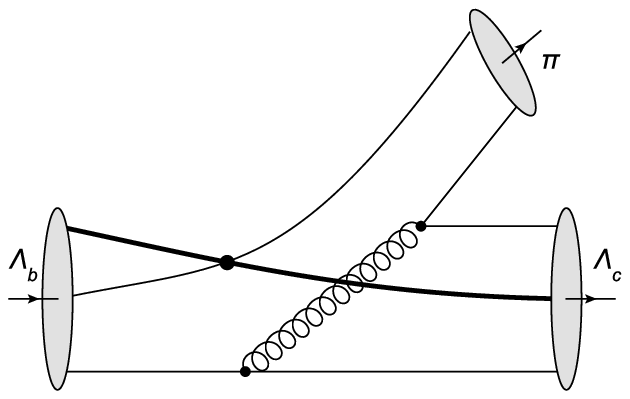}
\end{minipage}
}%
\hspace{0.04\linewidth}
\subfigure{
\begin{minipage}[b]{\minipagewidth}
\centering
\includegraphics[width=\subfigurewidth]{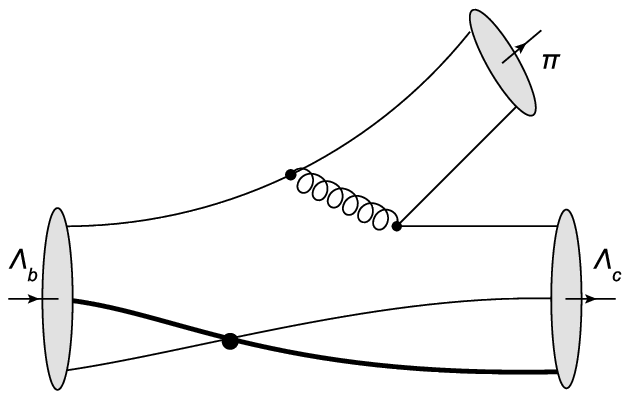}
\end{minipage}
}
\caption{\label{example2}Two other example diagrams that can be included in the diagrams in Fig. \ref{highfockstate} and should not be considered.}
\end{figure}

There are some diagrams that need not be considered at all.
Figure \ref{example1} is an example.
It looks like a correction to $T_B$ of the second kind at first sight.
However, noticing the crucial point that the gluon in this diagram is soft, it can be included in the wave function of $\Lambda_b$.
This means that this diagram can be absorbed into $T_B$.
In Fig. \ref{example2} we give two other examples.
Since the gluons are also soft, the two quarks generated form the gluon can be viewed as clouds in $\Lambda_b$.
This corresponds to $\Lambda_b$ in a nonvalence Fock state, \textit{i.e.}, $bduq\bar{q}$.
Then, the two diagrams in Fig. \ref{example2}, in fact, belong to the two diagrams in Fig. \ref{highfockstate}, respectively.
Now, we complete the proof of factorization for the decay $\Lambda_b^0\to\Lambda_c^+\pi^-$ at order $\alpha_s$.

The factorizable diagrams shown in Figs. \ref{tree1} and \ref{vertexcorrection} contribute to the decay amplitude of $\Lambda_b^0\to\Lambda_c^+\pi^-$ up to $\mathcal{O}(\alpha_s)$.
The decay amplitude is then
\begin{eqnarray}
\mathcal {A}_{\Lambda_b^0\to\Lambda_c^+\pi^-}
&=&\frac{G_F}{\sqrt{2}}V_{ud}^*V_{cb}\langle\pi^{-}|\bar{d}\gamma_{\mu}(1-\gamma_5)u|0\rangle \cdot\langle\Lambda_c|\bar{c}\gamma^{\mu}(a_{1V}-a_{1A}\gamma_5)b|\Lambda_b\rangle, \label{amplitude}
\end{eqnarray}
where
\begin{eqnarray}
a_{1j}
=\bar{c}_1(m_b)&+\frac{\bar{c}_2(m_b)}{N_c}
\bigg[1+\frac{\alpha_s(\mu)}{4\pi}C_F\int_0^1dx\Phi_{\pi}(x)F_{j}(x,z)\bigg], \label{a_1j}
\end{eqnarray}
with $j=V,A$.
The functions $F_{j}(x,z)$ are defined as
\begin{eqnarray}
F_{j}(x,z)=\left(3+2\ln\frac{x}{\bar{x}}\right)\ln z^2-7+f(x,\epsilon^jz)+f(\bar{x},\epsilon^j/z),
\end{eqnarray}
where $\bar{x}=1-x$, $\epsilon^j=1(-1)$ for $j=V(A)$, and the function $f$ has the following form:
\begin{widetext}
\begin{eqnarray}
f(x,z)&=&-\frac{x(1-z^2)[3+z-3x(1-z^2)]}{[1-x(1-z^2)]^2}\ln[x(1-z^2)-i\epsilon]-\frac{z}{1-x(1-z^2)}\nonumber\\*
&&+2\Big\{\Big[\frac{\ln[x(1-z^2)-i\epsilon]}{1-x(1-z^2)}-\ln^2[x(1-z^2)-i\epsilon]-\text{Li}_2[1-x(1-z^2)+i\epsilon]\Big]
-[x\to\bar{x}]\Big\}.\nonumber\\
\end{eqnarray}
\end{widetext}
In order to deduce Eq. (\ref{amplitude}), we have to deal with the loop integrals from the vertex-correction diagrams shown in Fig. \ref{vertexcorrection}.
Each of the diagrams in Fig. \ref{vertexcorrection} contains both ultraviolet and infrared divergences.
We isolate both of these kinds of divergences via dimensional regularization.
When neglecting the transverse momenta of the quarks, one can find that the infrared divergences of the four diagrams cancel.
Although this cancellation not happen when the transverse momenta are not neglected, the extra infrared divergence is still power-suppressed in the heavy-quark limit.
So, we can neglect the transverse momenta safely.
Then, we apply renormalization in the $\overline{\mathrm{MS}}$ scheme to get rid of the remaining ultraviolet divergence.

We can see that the coefficients $a_{1V}$ and $a_{1A}$ are independent of the scale and scheme at $\mathcal{O}(\alpha_s)$.
This is the same as the $B \to D\pi$ decay.
Although $a_{1V}$ and $a_{1A}$ gain scale dependence through the running coupling constant $\alpha_s(\mu)$, this dependence appears at higher orders in $\alpha_s$.

Noticing that the decay amplitude is Lorentz-invariant, Eq. (\ref{amplitude}) also holds in the rest frame of $\Lambda_b$, even though it is derived in the infinite-momentum frame of $\Lambda_b$.

With the values of the parameters from the Particle Data Group \cite{Nakamura:2010zzi}, we obtain the branching ratio for the decay $\Lambda_b^0\to\Lambda_c^+\pi^-$,
\begin{equation}
\mathrm{BR}(\Lambda_b^0\to\Lambda_c^+\pi^-)=|\zeta(\omega)|^2\times1.74\times10^{-2}\times(1\pm5.4\%),\label{numerical result}
\end{equation}
where the uncertainty (5.4\%) is mainly from the Cabibbo-Kobayashi-Maskawa matrix element $|V_{cb}|$.
If working in the naive-factorization approach, one would find the number $1.74$ in Eq. (\ref{numerical result}) is replaced by $1.64\pm0.06$, where the uncertainty comes mainly from the scheme and scale dependence.
The Isgur-Wise function $\zeta$ is model-dependent.
There are several models for the Isgur-Wise function: for example,
the soliton model \cite{Jenkins:1992se}, $\zeta(\omega)=0.99\mathrm{e}^{-1.3(\omega-1)}$; the MIT bag model \cite{Sadzikowski:1993iv}, $\zeta(\omega)=\left[2/(\omega+1)\right]^{3.5+\frac{1.2}{\omega}}$; and the Bethe-Salpeter-equation model \cite{Guo:1996jj}.
Our numerical results for the branching ratio are $(5.7\pm0.3)\times10^{-3}$, $(3.2\pm0.2)\times10^{-3}$, and $(4.5\pm0.9)\times10^{-3}$ in these three models, respectively, while the latest experimental data from the Particle Data Group is $(8.8\pm3.2)\times10^{-3}$ \cite{Nakamura:2010zzi}.
It can be seen that the soliton model and the Bethe-Salpeter equation model agree with the experimental data better.

In summary, we have proved that QCD factorization holds for $\Lambda_b^0 \to \Lambda_c^+ \pi^-$ and that the decay amplitude is renormalization-scale- and scheme-independent at $\mathcal{O}(\alpha_s)$ in the heavy-quark limit.
We find that, numerically, vertex-correction diagrams raise the branching ratio of $\Lambda_b^0 \to \Lambda_c^+\pi^-$ by about 6\% compared with that obtained in the naive-factorization approach.
Details of the proof will appear in our forthcoming paper \cite{ZhangGuo}.

\begin{acknowledgments}
This work is supported by the National Science Foundation of China (Projects Nos. 10675022 and 10975018) and the Fundamental Research Funds for the Central Universities (Project No. 2009SAP-3).
All the Feynman diagrams in this paper are drawn with the program JaxoDraw \cite{Binosi:2003yf}.
\end{acknowledgments}

\bibliography{zzh}

\newpage


\end{document}